\documentclass[preprint,prb,superscriptaddress,showpacs,aps,floatfix]{revtex4-2}%

\setcitestyle{maxcitenames=6}
\usepackage{graphicx}
\usepackage{color}
\usepackage{amsmath}
\usepackage{amsfonts}
\usepackage{subcaption}
\usepackage[percent]{overpic}
\usepackage{amssymb}
\usepackage{framed}
\usepackage{float}
\usepackage[normalem]{ulem}
\usepackage{lineno}
\usepackage{balance}
\providecommand{\U}[1]{\protect\rule{.1in}{.1in}}
\makeatletter
\renewcommand*{\fnum@figure}{{\normalfont\bfseries \figurename~\thefigure}}
\renewcommand*{\@caption@fignum@sep}{\normalfont\textbf{ : }}
\makeatother
\usepackage{hyperref}
\hypersetup{
    colorlinks=true,
    linkcolor=blue,
    filecolor=magenta,      
    urlcolor=cyan,
}
\newcommand{\erpbpd}{\text{ErPbPd}}

\begin{document}

\title{Anisotropy, frustration, and saddle point in the twisted kagome antiferromagnet ErPdPb}

\author{Resham Babu Regmi}
\email{Corresponding author: rregmi@nd.edu}
\affiliation{Department of Physics and Astronomy, University of Notre Dame, Notre Dame, IN 46556, USA}
\affiliation{Stavropoulos Center for Complex Quantum Matter, University of Notre Dame, Notre Dame, IN 46556, USA}

\author{Sk Jamaluddin}
\affiliation{Department of Physics and Astronomy, University of Notre Dame, Notre Dame, IN 46556, USA}
\affiliation{Stavropoulos Center for Complex Quantum Matter, University of Notre Dame, Notre Dame, IN 46556, USA}

\author{Y. Lee}
\affiliation{Ames National Laboratory, U.S. Department of Energy, Ames, Iowa 50011, USA}

\author{Hari Bhandari}
\affiliation{Department of Physics and Astronomy, University of Notre Dame, Notre Dame, IN 46556, USA}
\affiliation{Stavropoulos Center for Complex Quantum Matter, University of Notre Dame, Notre Dame, IN 46556, USA}
\author{Po-Hao Chang}
\affiliation{Department of Physics and Astronomy, George Mason University, Fairfax, VA 22030, USA}
\affiliation{Quantum Science and Engineering Center, George Mason University, Fairfax, VA 22030, USA}
\author{Peter E. Siegfried}
\affiliation{Department of Physics and Astronomy, George Mason University, Fairfax, VA 22030, USA}
\author{Abhijeet Nayak}
\affiliation{Department of Physics and Astronomy, University of Notre Dame, Notre Dame, IN 46556, USA}
\affiliation{Stavropoulos Center for Complex Quantum Matter, University of Notre Dame, Notre Dame, IN 46556, USA}

\author{Mohamed El Gazzah}
\affiliation{Department of Physics and Astronomy, University of Notre Dame, Notre Dame, IN 46556, USA}
\affiliation{Stavropoulos Center for Complex Quantum Matter, University of Notre Dame, Notre Dame, IN 46556, USA}

\author{Bence G. M\'arkus}
\affiliation{Department of Physics and Astronomy, University of Notre Dame, Notre Dame, IN 46556, USA}
\affiliation{Stavropoulos Center for Complex Quantum Matter, University of Notre Dame, Notre Dame, IN 46556, USA}

\author{Anna Ny\'ary}
\affiliation{Department of Physics and Astronomy, University of Notre Dame, Notre Dame, IN 46556, USA}
\affiliation{Stavropoulos Center for Complex Quantum Matter, University of Notre Dame, Notre Dame, IN 46556, USA}

\author{Zachary T. Messegee}
\affiliation{Department of Chemistry and Biochemistry, George Mason University, Fairfax, VA 22030, USA}

\author{Miya P. Zhao}  
\affiliation{The Chapin School, 100 East End Ave, New York, NY 10028}

\author{Xiaoyan Tan}
\affiliation{Quantum Science and Engineering Center, George Mason University, Fairfax, VA 22030, USA}
\affiliation{Department of Chemistry and Biochemistry, George Mason University, Fairfax, VA 22030, USA}

\author{L\'aszl\'o Forr\'o}
\affiliation{Department of Physics and Astronomy, University of Notre Dame, Notre Dame, IN 46556, USA}
\affiliation{Stavropoulos Center for Complex Quantum Matter, University of Notre Dame, Notre Dame, IN 46556, USA}

\author{Liqin Ke}
\affiliation{Department of Materials Science and Engineering, University of Virginia, Charlottesville, VA 22904}
\author{Igor I. Mazin}
\affiliation{Department of Physics and Astronomy, George Mason University, Fairfax, VA 22030, USA}
\affiliation{Quantum Science and Engineering Center, George Mason University, Fairfax, VA 22030, USA}
\author{Nirmal J. Ghimire}
\email{Corresponding author: nghimire@nd.edu}
\affiliation{Department of Physics and Astronomy, University of Notre Dame, Notre Dame, IN 46556, USA}
\affiliation{Stavropoulos Center for Complex Quantum Matter, University of Notre Dame, Notre Dame, IN 46556, USA}
\date{\today}

\begin{abstract}
  The kagome lattice, with its inherent geometric frustration, provides a rich platform for exploring intriguing magnetic phenomena and topological electronic structures. In reduced-symmetry structures, such as twisted kagome systems involving rare earth elements, additional anisotropy can arise, enabling intriguing properties including spin-ice states, magnetocaloric effects, noncollinear magnetic ordering, and anomalous Hall effect. Here, we report the synthesis of single crystals of ErPdPb, which features a twisted kagome lattice net of Er atoms within the hexagonal ZrNiAl-type structure, and we investigate its magnetic, electronic, and thermal properties. The material exhibits antiferromagnetic ordering below 2.2 K, consistently observed in magnetic, transport, and heat capacity measurements. Magnetization measurements reveal 1/3 metamagnetic steps along the $c$-axis below the N\'eel temperature, suggesting an Ising-spin-like state on the twisted kagome lattice. A pronounced anisotropy between in-plane and out-of-plane resistivity is observed throughout the temperature range of 1.8–300 K, and the compound exhibits a significant frustration index of 13.6 (12.7) along the $c$-axis ($ab$-plane). Heat capacity measurements show a broad hump at 2.2 K, with an additional increase below 0.5 K. The anisotropic magnetic properties are further explored through density functional theory (DFT) calculations, which suggest  strong easy-axis anisotropy, consistent with experimental magnetic measurements and crystal-field model expectations, and quasi-one-dimensional bands and a spin-split saddle point at the zone center.

\end{abstract} 
\maketitle

\section{Introduction}\label{sec:1}
\begin{figure}[ht!]
\centering
\includegraphics[width=0.8\columnwidth, keepaspectratio, scale=1]{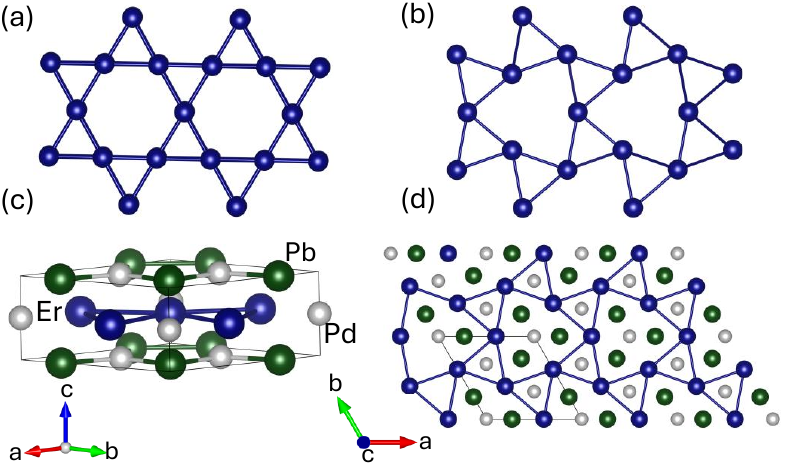}
\caption{\small \textbf{Crystal structures of Kagome and Twisted Kagome system.} (a) A planar prototypical kagome arrangement with  corner sharing equilateral triangles forming a kagome structure in the ab-plane. In contrast, Figure (b) emphasizes a crucial structural variation, illustrating the twisted kagome net in the $ab$ plane
(c) depicts the crystal structure of ErPdPb in 3D (d) $c$-axis view of the crystal structure emphasizing the twisted kagome net of Er atoms in the $ab$-plane.}
\label{Fig1}
\end{figure}

The kagome lattice, in which atoms are arranged in corner-sharing equilateral triangles as shown in Fig. \ref{Fig1}(a), is known to exhibit strong geometric frustration in two dimensions \cite{ghimire2020topology}. This frustration gives rise to a variety of non-trivial quantum states, including classical and quantum spin liquids \cite{RevModPhys.89.025003,doi:10.1126/science.aay0668,balents2010spin,faak2012kapellasite,iqbal2015paramagnetism,barthelemy2019local,hering2022phase,feng2017gapped} and spin-ice states \cite{castelnovo2008magnetic}. Kagome lattice systems also host non-collinear spin structures \cite{huang2023crrhas,ghimire2020topology} and intriguing electronic features such as flat bands, Dirac points, and both integer and fractional quantum Hall states \cite{tang2011high,neupert2011fractional,xu2015intrinsic}, all of which have attracted significant research interest.

A closely related structural motif is the twisted kagome lattice, in which the equilateral triangles are buckled as depicted in Fig. \ref{Fig1}(b) \cite{huang2023crrhas, bhandari2025tunable}. The ZrNiAl-type structure is a prototypical example, featuring a twisted kagome network of Zr atoms. In this structure, the buckling of Zr atoms breaks inversion symmetry, resulting in a hexagonal lattice with space group $P\bar{6}2m$. This structure-type is known for its extraordinary chemical versatility, incorporating compounds with all lanthanides, elements from groups 3A, 4A, and 5A, and even certain actinides, such as uranium \cite{Villars2023-24, huang2023crrhas}.

The RAgGe compounds, which crystallize in the ZrNiAl-type structure, have attracted considerable attention for their rich magnetic and electronic properties \cite{GIBSON199634, morosan2004thermodynamic, morosan2005angular, baran1998magnetic}. Recently, HoAgGe has been highlighted as a promising candidate for realizing a two-dimensional (2D) spin-ice model, attributed to the strong in-plane anisotropy of Ho spins \cite{zhao2020realization, li2022low, wu2025lattice}. In addition, these materials exhibit intriguing band topology and unconventional magnetotransport properties \cite{bhandari2025tunable}.

Within this broader family, the lead-based members, particularly the $R$PdPb compounds, have been synthesized and structurally characterized, primarily in polycrystalline form \cite{marazza1995contribution, iandelli1994structure}; however, their single-crystal growth and physical properties remain largely unexplored. The incorporation of the heavy Pb atom introduces substantially stronger spin–orbit coupling compared to the lighter Ge analogs in $R$AgGe, which can significantly entangle spin and orbital degrees of freedom. This enhanced spin–orbit interaction is expected to play a decisive role in determining the magnetic anisotropy and electronic structure, potentially giving rise to pronounced magnetocrystalline anisotropy, anisotropic magnetotransport, and possibly topological behavior.

Here, we report crystal growth and investigate the physical and electronic properties of ErPdPb, which crystallizes in the ZrNiAl-type structure as shown in Fig. \ref{Fig1}(c), and in which Er atoms form the twisted kagome net as depicted in Fig. \ref{Fig1}(d). This compound undergoes antiferromagnetic ordering below 2.2 K, exhibiting easy-axis magnetic anisotropy. It shows metallic behavior over the full temperature range from 300 K to 1.8 K. The in-plane resistivity is significantly higher than the out-of-plane ($c$-axis) resistivity, a characteristic feature also observed in the $R$AgGe family of compounds \cite{bhandari2025tunable}. Interestingly, heat capacity measurements reveal a broad hump near the antiferromagnetic transition, followed by a sharp upturn below 0.5 K to 50 mK, suggesting the possibility of an additional magnetic transition below 50 mK. Furthermore, magnetization (magnetic moment vs. magnetic field) measurements show a 1/3 plateau, marking ErPdPb as a materials platform for exploring unusual low-temperature magnetic phenomena. If a magnetic transition does indeed occur below 50 mK, further suppression of the ordering temperature via physical or chemical pressure could enhance quantum fluctuations, potentially giving rise to the emergent phenomena such as superconductivity or non-Fermi liquid behavior. DFT calculations show that ErPdPb exhibits strong easy-axis anisotropy, consistent with experimental observations and predict contrasting anisotropy trends between Ho and Er. Band structure calculations show the presence of saddle point, a characteristic feature of the kagome geometry, close to the Fermi energy.

\section{Materials and Methods}

\begin{figure}[ht!]
\centering
\includegraphics[width=0.8\columnwidth, keepaspectratio, scale=1]{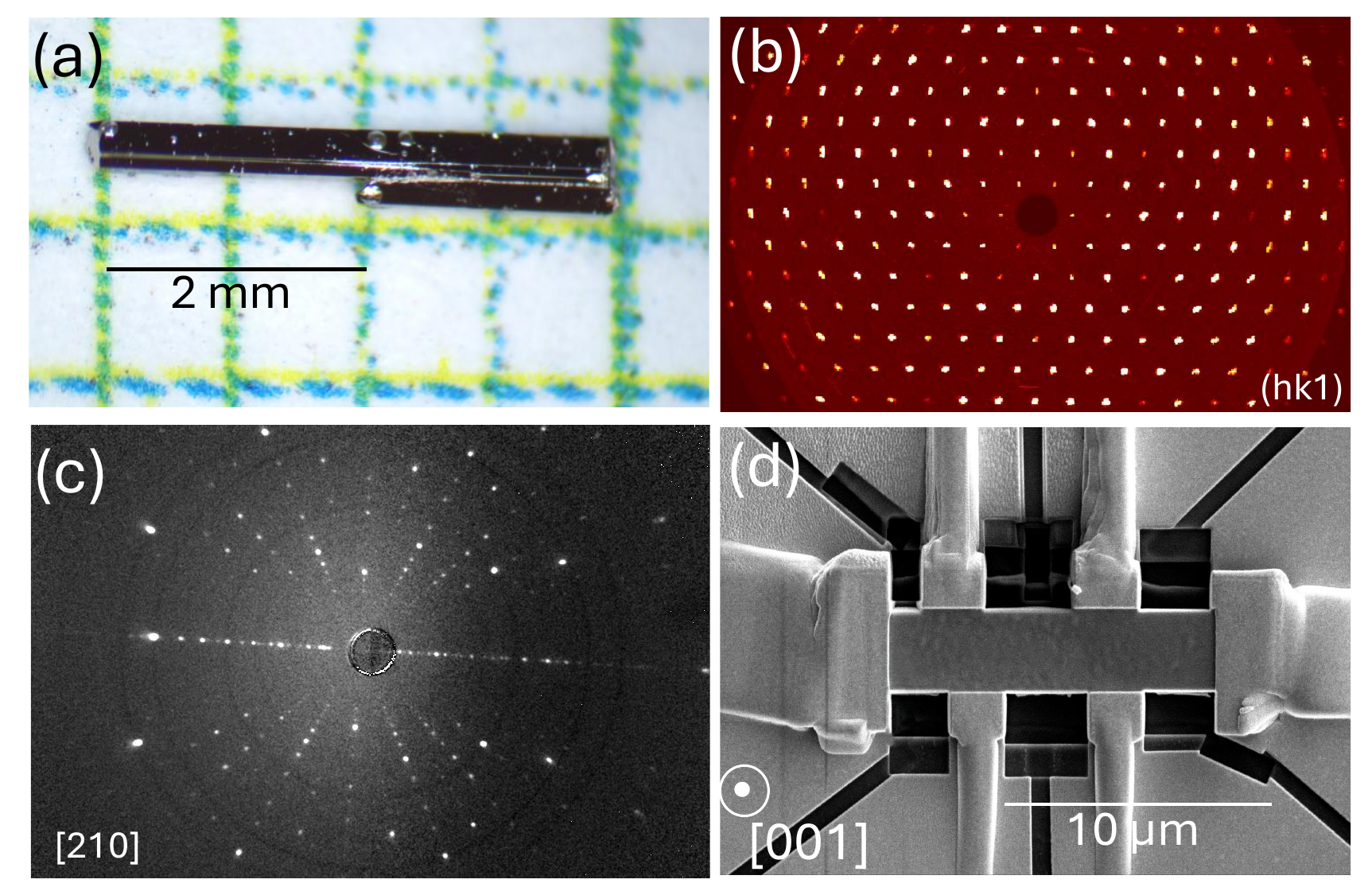}
    \caption{\small \textbf{Optical image, single crystal precession image and FIB device of ErPdPb.} (a) Optical image of a needle like crystal (b) Single crystal XRD precession image of \textit{(hk1)} plane showing a hexagonal symmetry of ErPdPb (c) Laue pattern on single crystal of ErPdPb surface along a [210] direction (d) FIB fabricated Hall bar with [001] axis normal to the surface.}
    \label{single crystal}
\end{figure}
\textbf{Single crystal growth.} Single crystals of ErPdPb were grown using the self-flux method with Pb as the flux. High-purity elements—Er (Alfa Aesar, 99.9\%), Pd (Alfa Aesar, 99.9\%), and Pb (Alfa Aesar, 99.9\%)—were loaded in a 2 mL aluminum oxide crucible in a molar ratio of 1:1:10. The crucible was then sealed in a fused silica ampoule under vacuum. The ampoule was heated to 1150$^{\circ}$C over 12 hours, held at that temperature for another 12 hours, and subsequently cooled to 700 $^{o}$C over 100 hours. Once the furnace reached 700$^{\circ}$C, the ampoule was removed and decanted using a centrifuge to separate the Pb flux. Needle-like single crystals of several mm in length and weighing approximately 5–7 mg as depicted in Fig. \ref{single crystal}(a) were obtained. These crystals often contained residual lead flux on their surfaces, which was mechanically removed using either a razor blade or fine abrasive films.

\textbf{Single crystal X-ray diffraction.} Structural characterization was performed using single-crystal X-ray diffraction (XRD) at room temperature. Data were collected on a crystal with dimensions of $0.12\times0.08\times0.08$ mm$^{3}$, using a Rigaku XtaLAB Synergy-i diffractometer with the HyPix-Bantam direct photon-counting detector. Single crystal was mounted on a loop and measured on the goniometer head of the diffractometer. Reduction of X-ray diffraction intensities was carried out using the CrysAlisPro v 42.55a package. The crystal structures were both solved with space group $P\overline{6}2m$ and refined using SHELX-2018 software \cite{sheldrick2008short}. The refined structure parameters are summarized and presented in Table \ref{T1}. Corresponding fractional positions and anisotropic thermal parameters are also provided (Supporting Information, Table I). For reproducibility singl crystal X-ray diffraction was also collected in another sample by using a Bruker Quest diffractometer equipped with a Bruker PHOTON III detector. A measured ($hk1$) zone obtained on Bruker Quest diffractometer of ErPdPb single crystal is shown in Fig. \ref{single crystal}(b). The crystal orientation was confirmed prior to each measurement using a Photonic Science X-ray Laue Crystal Orientation System. A representative Laue diffraction pattern collected along [210] direction is shown in Fig. \ref{single crystal}(c).

\textbf{Chemical analysis.} The elemental composition was analyzed on the surface of a single crystal using an Octane Elect Plus energy-dispersive X-ray spectroscopy (EDX) system, integrated with a JEOL JSM-IT500HRLV scanning electron microscope (SEM). Elemental maps were acquired at an accelerating voltage of 15 keV.

\textbf{Magnetic measurements.} Magnetic properties were measured using a 7-T Quantum Design Magnetic Property Measurement System (MPMS3) over a temperature range of 1.8 to 400 K using a VSM mode with a peak amplitude of 8 mm and averaging time of 2 seconds . A thin, needle-like single crystal was mounted on a quartz holder using GE varnish. Magnetization and magnetic susceptibility were measured along both the $ab$-plane and the $c$-axis after confirming the crystal orientation.

\textbf{Electrical transport measurements.} Electrical resistivity, magnetoresistance, and Hall resistivity were measured using a 9-T Quantum Design Dynacool Physical Property Measurement System (QD PPMS) from 1.8  to 300 K. For measurements with electrical current along the $c$-axis (along the needle length), a single crystal was trimmed into a Hall bar geometry. Standard 4-probe measurements were performed using 25 $\mu$m Pt wires (metal basis, 99.95 \%) attached with Epotek H20E silver epoxy, cured at 120 $^{\circ}$C. Transport measurements were conducted with an excitation current of 2 mA. Magnetoresistance (MR), was calculated as MR = ($\rho_{xx}(B)$- $\rho_{xx}(B= 0)$)/$\rho_{xx}(B= 0)$. For measurement with $B||c$, $\rho_{xx}$ = $\rho_{ab}$, the MR is labeled MR$_c$, and for the $B||ab$-plane, $\rho_{xx}$ = $\rho_{c}$, the MR is labeled MR$_{ab}$. 

To correct for contact misalignment, the magnetoresistance data were symmetrized, following the procedures described in \cite{bhandari2025tunable}.Due to the small diameter of the needle-shaped crystals (less than 0.5 mm), in-plane ($ab$-plane) electrical transport measurements required nanofabrication. A Hall bar sample was prepared using a Ga$^+$-based focused ion beam (FIB) as shown in Fig. \ref{single crystal}(d). The longitudinal contact leads had a length of 2.833 micrometers ($\,\mu\text{m}$)
, a width of 2.564 $\,\mu\text{m}$, and a thickness of 860 nanometers ($\,\text{nm}$).

\textbf{Heat capacity measurements.} Heat capacity between 1.8 and 200 K was measured using the standard option in the QD PPMS. The low temperature heat capacity (between 50 mK and 4 K) was measured using a QD PPMS equipped with a $^{3}\mathrm{He}/^{4}\mathrm{He}$ dilution refrigerator. A 2.2 mg single crystal was used for these measurements.

\textbf{Density functional theory calculations.} 
The DFT calculations were performed using the full-potential linear augmented plane wave (FP-LAPW) method, as implemented in \textsc{Wien2k}~\cite{WIEN2k}.
The generalized gradient approximation of Perdew, Burke, and Ernzerhof~\cite{perdew1996prl} was used for the correlation and exchange potentials.
Spin-orbit coupling (SOC) was included using the second variational method.
To generate the self-consistent potential and charge, we employed $R_\text{MT}\cdot K_\text{max}=9.0$ with muffin-tin (MT) radii $R_\text{MT}=$ 2.8, 2.5, and 2.5 a.u.~for all Er, Pb, and Pd atoms, respectively.
Crystal parameters were based on experimental data from this study.
The calculations are performed with 6300 $k$-points in the Full Brillouin zone (FBZ) and iterated until the charge differences between consecutive iterations were smaller than $10^{-5} e$ and the total energy differences lower than $10^{-3}$ mRy/cell.
The strongly correlated Er-$4f$ electrons were treated using the DFT+$U$ method with the fully-localized-limit (FLL) double-counting scheme~\cite{anisimov1993prb}.
We employed multiple $U$-values to understand the effects of $U$ on magnetic properties of $\erpbpd$.
Importantly, the Er-$4f$ states were constrained to satisfy Hund's rules\cite{lee2025ncm}.

The magnetocrystalline anisotropy (MA) was evaluated by calculating total energy differences as a function of the spin quantization direction.
Rare-earth MA is among the most challenging intrinsic magnetic properties to capture within DFT.
For heavy rare-earths such as Er, the fully polarized Hund’s rule state can be represented by a single Slater determinant in DFT-based approaches~\cite{ke2025a}.
However, the orbital dependence of self-interaction errors and the lack of orbital polarization typically drive the Hund's rule ground state into metastability unless explicit constraints are applied. 
The erroneous $4f$ orbital occupancy results in a wrong 4f charge distribution, which translates into an incorrect MAE.
Recently, we have demonstrated that this can be efficiently resolved by enforcing Hund's rules using constrained DFT~\cite{lee2025ncm}.
Benchmarking on several dozen compounds has shown that the MA calculated using this approach agrees well with low-temperature experiments~\cite{riberolles2022prx,rosenberg2022prb,lee2023prb,xu2024jacs,han2024a,lee2025ncm,gazzah2025arXiv}.

\section{Results and Discussion}\label{sec:4}

\subsection{Crystal Structure}

\begin{table}[h]
\caption{Structure refinement parameters for the single crystal of ErPdPb. Numbers in parentheses are estimated standard deviation.}\label{T1} 
\begin{tabular}
 [c]{c@{\hspace{0.3cm}}c@{\hspace{0.3cm}}c@{\hspace{0.3cm}}c@{\hspace{0.3cm}}c@{\hspace{0.3cm}}c}\hline   

Crystal system & Hexagonal (ZrNiAl-type) \\
Space group & $P\overline{6}2m$  \\
Temperature (K) & 293(2)  \\
Wavelength (\AA) & 0.71073  \\
Z formula units & 3       \\
crystal size ($\text{mm}^3$) &	0.12 x 0.08 x 0.08 \\
2$\theta_\textrm{min}$ & 5$^\circ$  \\
2$\theta_\textrm{max}$ & 30.500$^\circ$           \\ 
Formula weight (g/mol) & 480.88  \\
$a,b,c$ (\AA) & 7.70710(1), 7.70710(1), 3.78200(1)\\
Volume (\AA)$^3$ &194.551(7)             \\
Density (calculated) (g/cm$^3$) & 12.313                \\
unique reflections & 212              \\
parameters refined & 14             \\
 $\mu$ (Mo $K_{\alpha}$) cm$^{-1}$ & 103.258              \\
Goodness-of-fit on $F^2$ & 1.128  \\
$\mathrm{R_1,\ wR_2\ [F_o > 4\sigma(F_o)]}$ & $\mathrm{0.0177,\ 0.0477}$ \\

\end{tabular}
\par%
\begin{tabular} 
[c]{c@{\hspace{0.3cm}}c@{\hspace{0.3cm}}c@{\hspace{0.3cm}}c@{\hspace{0.3cm}}c@{\hspace{0.3cm}}c@{\hspace{0.3cm}}c}\hline

 Atom & Wyck. & Occ. & x & y & z     \\ \hline
Er   &   3g  & 1 &   0.60242(1)   &   0.000000   & 0.500000  \\
Pd1  &    1b & 1 &  0.000000   &   0.000000   & 0.500000 \\
Pd2    &  2c   & 1 &   0.666667   & 0.333333      & 0.000000 \\
Pb    &  3f   & 1 & 0.26842(1)     &  0.000000  & 0.000000 \\ 

 \hline

\end{tabular}
\end{table}

ErPdPb was first synthesized in polycrystalline form by A. Iandelli \cite{iandelli1994structure} and R. Marazza et al. \cite{marazza1995contribution} in the 1990s. Its crystal structure was determined using powder X-ray diffraction and identified as the Fe$_2$P-type (or ZrNiAl-type) structure, with reported lattice parameters of a = 7.704, \AA and c = 3.794 \AA \cite{marazza1995contribution}. To the best of our knowledge, single crystals of ErPdPb had not been previously grown, and its structural information was solely based on powder X-ray diffraction data. In this work, we performed single-crystal X-ray diffraction on the crystals we synthesized, which confirmed the structure of ZrNiAl with the hexagonal space group $P\overline{6}2m$. This structure includes a twisted kagome net formed by Er atoms. Details of structural refinement parameters are provided in Table \ref{T1}, and a schematic illustration of the crystal structure is shown in Figs. \ref{Fig1}(c) and \ref{Fig1}(d).

\subsection{Magnetic Susceptibility and Magnetization}

\begin{figure*}[ht!]
\begin{center}
\includegraphics[width=.8\linewidth]{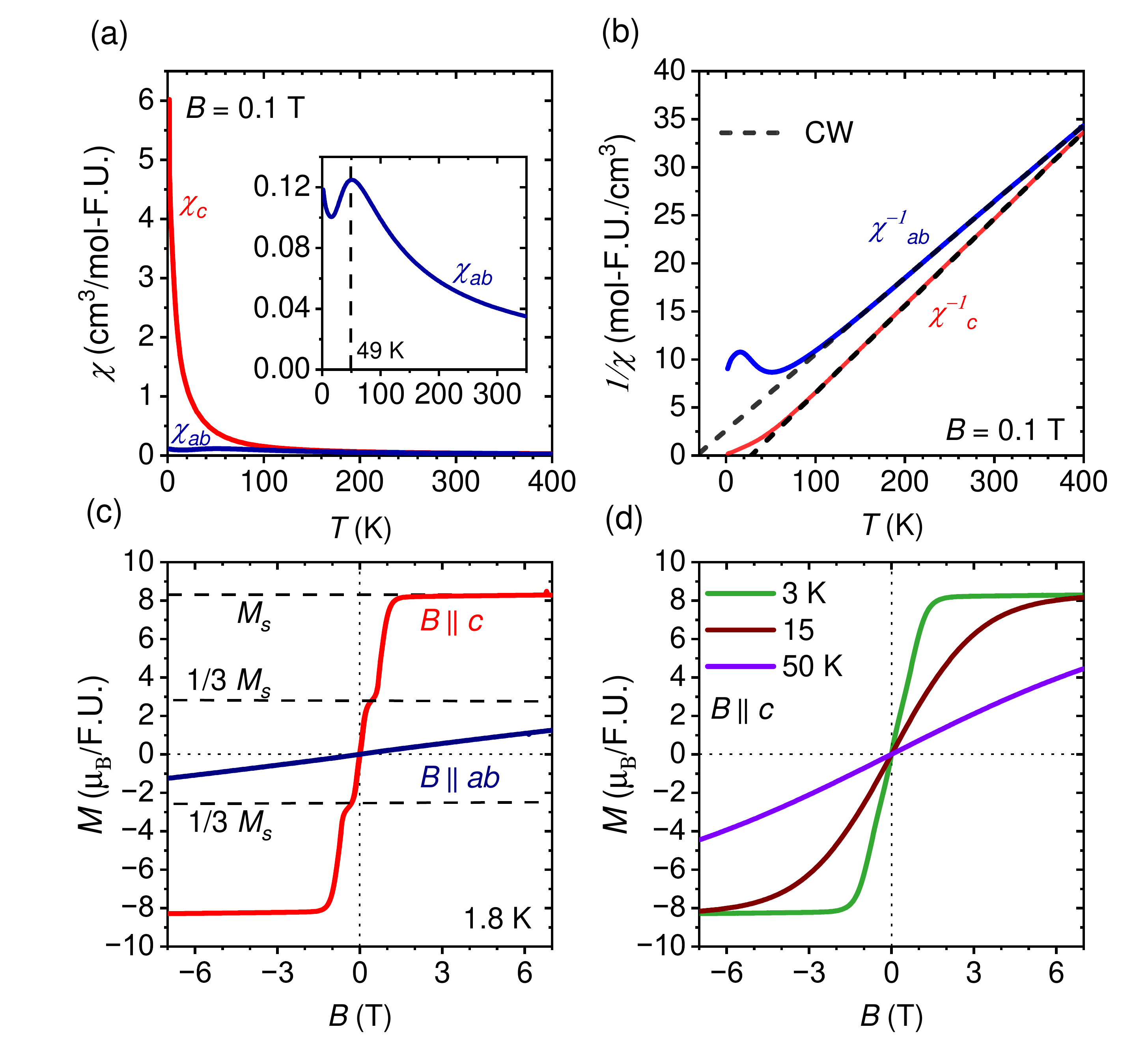}
    \caption{\small \textbf{Magnetic properties of ErPdPb.} (a) Magnetic susceptibility measured in a magnetic field ($B=\mu_{0}H$) of 0.1 T applied along crystallographic $c$-axis ($\chi_c$) and within the $ab$-plane ($\chi_{ab}$) measured with a field-cooled (FC) protocol. Inset highlights a broad feature centered at 49 K for $\chi_{ab}$ (b).  Inverse susceptibility of ErPdPb for $B$ along $c$-axis  ($\chi^{-1}_{c}$) and along $ab$-plane ($\chi^{-1}_{ab}$). The dashed lines represent the Curie-Weiss (CW) fit to the data. (c) Magnetic moment for $B||c$-axis (red curve), and $ab$-plane (blue curve) measured at 1.8 K. (d) Magnetic moment for $B||c$-axis measured at 3, 15 and 50 K.} 
      \label{Fig3}
    \end{center}
\end{figure*}

Magnetic measurements are presented in Fig. \ref{Fig3}. Fig \ref{Fig3}(a) shows the magnetic susceptibility ($\chi$) as a function of temperature performed on an oriented single crystal of ErPdPb in the temperature range between 400 and 1.8 K, measured with an external magnetic field of 0.1 T applied along the $c$-axis ($\chi_{c}$) and within the $ab$-plane ($\chi_{ab}$). $\chi_{ab}$ shows a small variation with temperature. However, $\chi_{c}$ increases significantly below 100 K. A significant cusp-like feature is observed in $\chi_{ab}$ centered at 49 K, as highlighted in the inset of Fig. \ref{Fig3}(a). This feature usually indicates either low-dimensional correlations or short-range magnetic order. Given the two-dimensional (2D) nature of Er atoms and the enhanced in-plane anisotropy due to the twisting observed in the isostructural HoAgGe \cite{zhao2020realization, bhandari2025tunable}, either of these reasons is possible. Local probe methods such as muon spin rotation or neutron pair distribution function (PDF) studies are required to pinpoint the exact origin. There are no other features in the susceptibility in either direction, representing a typical magnetic ordering behavior. Although a sharp magnetic transition is not clearly observed in the susceptibility data alone, measurements of magnetization (magnetic moment $M$ vs. magnetic field $B$), resistivity, and heat capacity (discussed below) consistently indicate antiferromagnetic ordering below the Néel temperature ($T_{\mathrm{N}}$) of 2.2 K.

The inverse magnetic susceptibility, $\chi^{-1}$, for both crystallographic directions is shown in Fig. \ref{Fig3}(b). At high temperatures, both $\chi^{-1}_{c}$ and $\chi^{-1}_{ab}$ exhibit linear behavior, indicative of Curie-Weiss paramagnetism. A Curie-Weiss fit of the form $\chi = C/(T - \theta_{\mathrm{CW}})$, where $C = N_A\mu_{\mathrm{eff}}^2/3k_B$ is the Curie constant, $N_A$ is the Avogadro number and $k_B$ is the Boltzmann constant, produces an effective magnetic moment $\mu_{\mathrm{eff}}$ of 9.4 $\mu_B$ and a Curie-Weiss temperature $\theta_{\mathrm{CW}}$ of 28.2 K for $\chi^{-1}{c}$. Similarly, the fit to $\chi^{-1}{ab}$ gives $\mu_{\mathrm{eff}} = 10.0$ $\mu_B$ and $\theta_{\mathrm{CW}} = -30$ K. In both directions, the experimentally extracted values of $\mu_{\mathrm{eff}}$ are close to the theoretical value of 9.6 $\mu_B$ expected for an Er$^{3+}$ ion. The large anisotropy of $\theta_{\mathrm{CW}}$
suggests that the anisotropic exchange terms, most likely the Ising exchange, are very strong and of opposite sign to the isotropic Heisenberg exchange, likely due to the presence of heavy Pb ions between the Er layers. The large frustration index $f = \theta_{\mathrm{CW}}/T_{\mathrm{N}}$, of 13.6 ($c$-axis) and 12.7 ($ab$-plane) indicates a strong role of 2D Mermin-Wagner fluctuations suppressing the 3D ordering.

Magnetization ($MB$) measurements, presented in Figs. \ref{Fig3}(c) and \ref{Fig3}(d), provide further insight into the nature of magnetic ordering in this material. Figure \ref{Fig3}(c) shows the $MB$ data measured with $B$ between $\pm$ 7 T applied along the $c$-axis ($MB_{c}$, red curve) and on the $ab$-plane ($MB_{ab}$, blue curve) at 1.8 K. The curve $MB_{ab}$ exhibits linear behavior, while $MB_{c}$ shows the saturated moment at \( M_s = 8.3\,\mu_B/\mathrm{F.U.} \)
, a characteristic signature of the easy-axis ($c$-axis) antiferromagnetic ordering within the equilateral triangles of a kagome lattice. The $MB_{ab}$ response remains linear at all measured temperatures (data not shown). However, $MB_{c}$ behavior evolves significantly with temperature. As shown in Fig. \ref{Fig3}(d), by 3 K the 1/3-magnetization (1/3 $M_s$) plateau disappears, and the curve shows only a monotonic approach to magnetic saturation. This behavior persists even at 15 K, and some remnants of it are still visible at 50 K. These observations suggest the presence of short-range ferromagnetic correlations polarized along the $c$-axis at temperatures well above the Néel temperature, consistent with strong two-dimensional magnetic fluctuations in the twisted kagome net, which suppress long-range order while preserving substantial local correlations.
 
\subsection{Resistivity, Magnetoresistance and Heat Capacity}

 \begin{figure*}[ht!]
\centering
\includegraphics[width=1\linewidth]{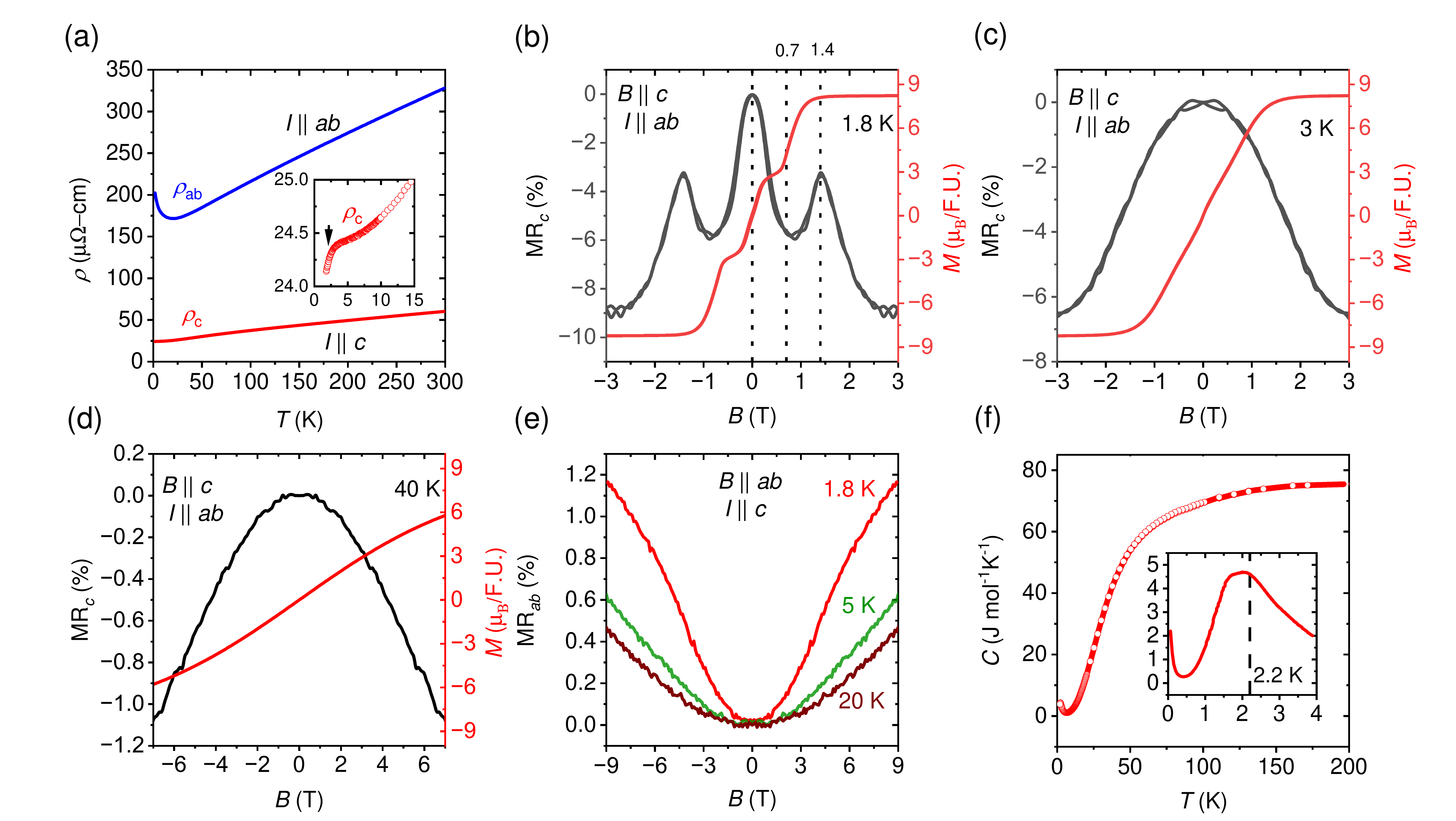}
    \caption{\small \textbf{Physical properties of ErPdPb.} (a) Zero field Electrical resistivity as a function of temperature with $I$ $||$ $c$-axis ($\rho_{c}$) and \textit{I} $||$ \textit{ab}-plane ($\rho_{ab}$). (b) Longitudinal Magnetoresistance for \textit{I} $||$ $ab$-plane with $B$ $||$ $c$-axis and (black curve) magnetization with field along same direction (red curve) overlayed on a single plot at 1.8 K. (c) Magnetoresistance measured with $B$ $||$ $ab$-plane and $I$ $||$ $c$-axis and  magnetization with field along same direction overlayed on a same plot at 3 K. (d) Longitudinal magnetoresistance for \textit{I} $||$ $ab$-plane and \textit{B} $||$ $c$ at 40 K. (e) Longitudinal magnetoresistance for \textit{I} $||$ $c$-axis and $B$ $||$ $ab$ plane. (f) Heat Capacity as a function of temperature with inset showing low temperature range 0.5- 4 K.  Inset clearly shows the broad anomaly at 2.2 K and sharp upturn below 0.5 K.} 
    \label{Fig4}
   
\end{figure*}

Transport and thermal data are presented in Fig. \ref{Fig4}. The zero-field resistivity, $\rho$, as a function of temperature is shown in Fig. \ref{Fig4}(a). The red curve corresponds to the resistivity measured with electric current applied along the $c$-axis ($\rho_{c}$), while the blue curve corresponds to current applied within the $ab$-plane ($\rho_{ab}$). Across the entire temperature range, $\rho_{c}$ is consistently lower than $\rho_{ab}$. This behavior is typical for twisted kagome compounds with a ZrNiAl-type structure, such as HoAgGe \cite{bhandari2025tunable}, and is attributed to the anisotropic Fermi surface. The low-temperature behavior of $\rho_{c}$ is magnified in the inset of Fig. \ref{Fig4}(a), where a distinct drop in resistivity is observed at 2.2 K. This drop reflects a reduction in spin scattering due to antiferromagnetic ordering below this temperature, marking 2.2 K as the Néel temperature, $T_{\mathrm{N}}$. $\rho_{c}$ exhibits metallic behavior across the full temperature range. In contrast, $\rho_{ab}$ decreases with decreasing temperature from 300 K to approximately 25 K, below which it increases, showing no noticeable anomaly at $T_{\mathrm{N}}$. This upturn may originate from crystal field effects and has also been observed in HoAgGe \cite{bhandari2025tunable}. 

Although $\rho_{ab}$ does not exhibit any anomaly at $T_{\mathrm{N}}$, the magnetoresistance MR$_{c}$ (measured with current in the $ab$-plane and magnetic field applied along the $c$-axis) displays a distinct feature associated with the 1/3-magnetization step in $MB_{c}$. MR$_{c}$ and $MB_{c}$ measured at 1.8 K are shown together in Fig. \ref{Fig4}(b). The slight mismatch between the magnetic field at which the 1/3-magnetization step appears in $MB_{c}$ and the sharp change in MR$_{c}$ can be attributed to the demagnetization effect. This arises because MR was measured on a very thin sample fabricated using focused ion beam (FIB), whereas magnetization was measured on a larger bulk sample. At 3 K (Fig. \ref{Fig4}(c)), MR$_{c}$ is smooth and negative, leveling off above ~2.5 T, resembling the magnetization curve that saturates beyond 1.5 T. MR$_{c}$ remains negative even at 40 K (Fig. \ref{Fig4}(d)). In contrast, MR$_{ab}$, shown in Fig. \ref{Fig4}(e), is positive across all measured temperatures from 1.8 K to 20 K. These observations provide magnetotransport evidence for easy-axis antiferromagnetic ordering below 2.2 K in ErPdPb.

\begin{figure}[ht!]
   \begin{tabular}{c}
    \includegraphics[width=0.5\linewidth,clip]{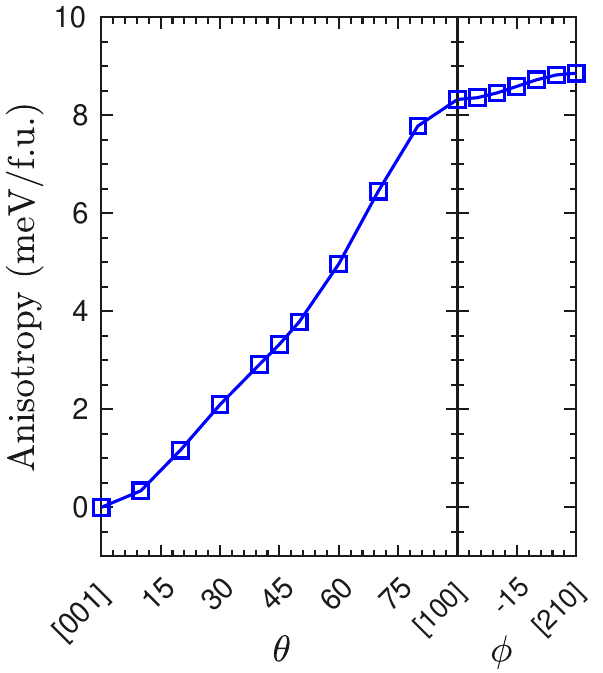}
  \end{tabular}%
  \caption{Magnetocrystalline anisotropy energy (in meV/f.u.) in ErPdPb, represented by the variation of magnetic energy as a function of spin-axis rotation, calculated using DFT+$U$, with $U$=10~eV applied on Er-$4f$ orbitals.
}
  \label{fig:mae}
\end{figure}

The heat capacity of ErPdPb is shown in Fig. \ref{Fig4}(f). Initial measurements were conducted from 200 K down to 1.8 K, and the resulting data are presented in the main panel. Subsequently, the same sample was measured using a dilution refrigerator in the range of 4 K to 50 mK, as shown in the inset of Fig. \ref{Fig4}(f). The heat capacity decreases steadily to about 10 K, below which it increases, reaching a broad peak at approximately 2.2 K. Below this temperature, it drops sharply, forming a broad lambda-like feature. Interestingly, below 0.5 K, the heat capacity begins to rise again. This low temperature upturn could originate from various sources, such as the Kondo effect or the onset of magnetic ordering at very low temperatures. Further investigation is necessary to determine whether quantum fluctuations can be enhanced by chemical or physical pressure, resulting in emergent properties such as non-Fermi liquid behavior and superconductivity.
  
\subsection{First principles calculations}

\begin{figure*}[ht!]
\includegraphics[width=.37\linewidth]{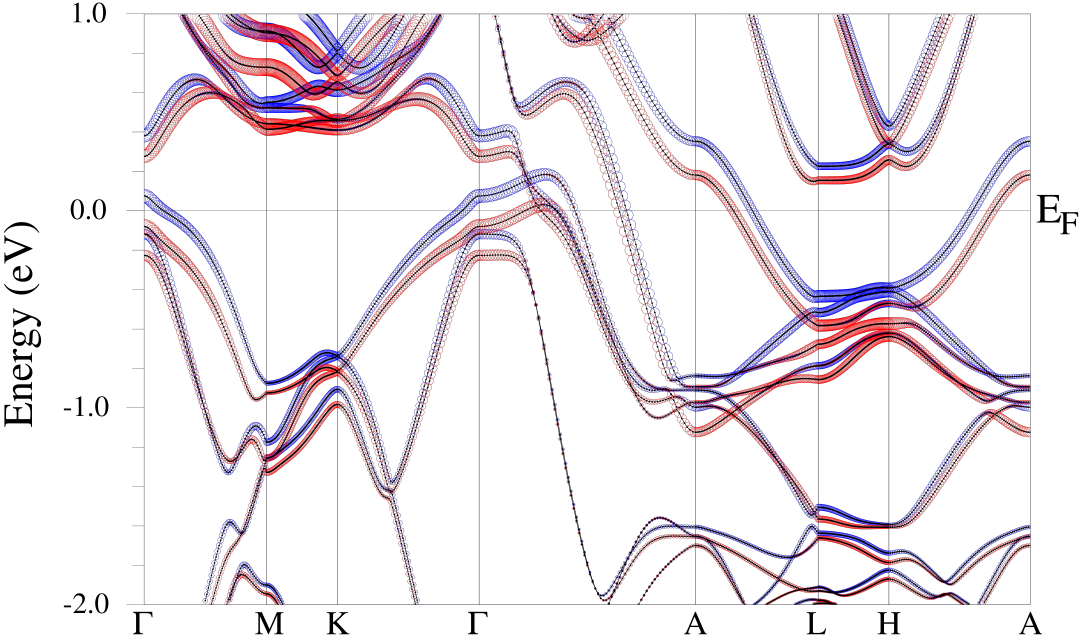}
\includegraphics[width=.16\linewidth]{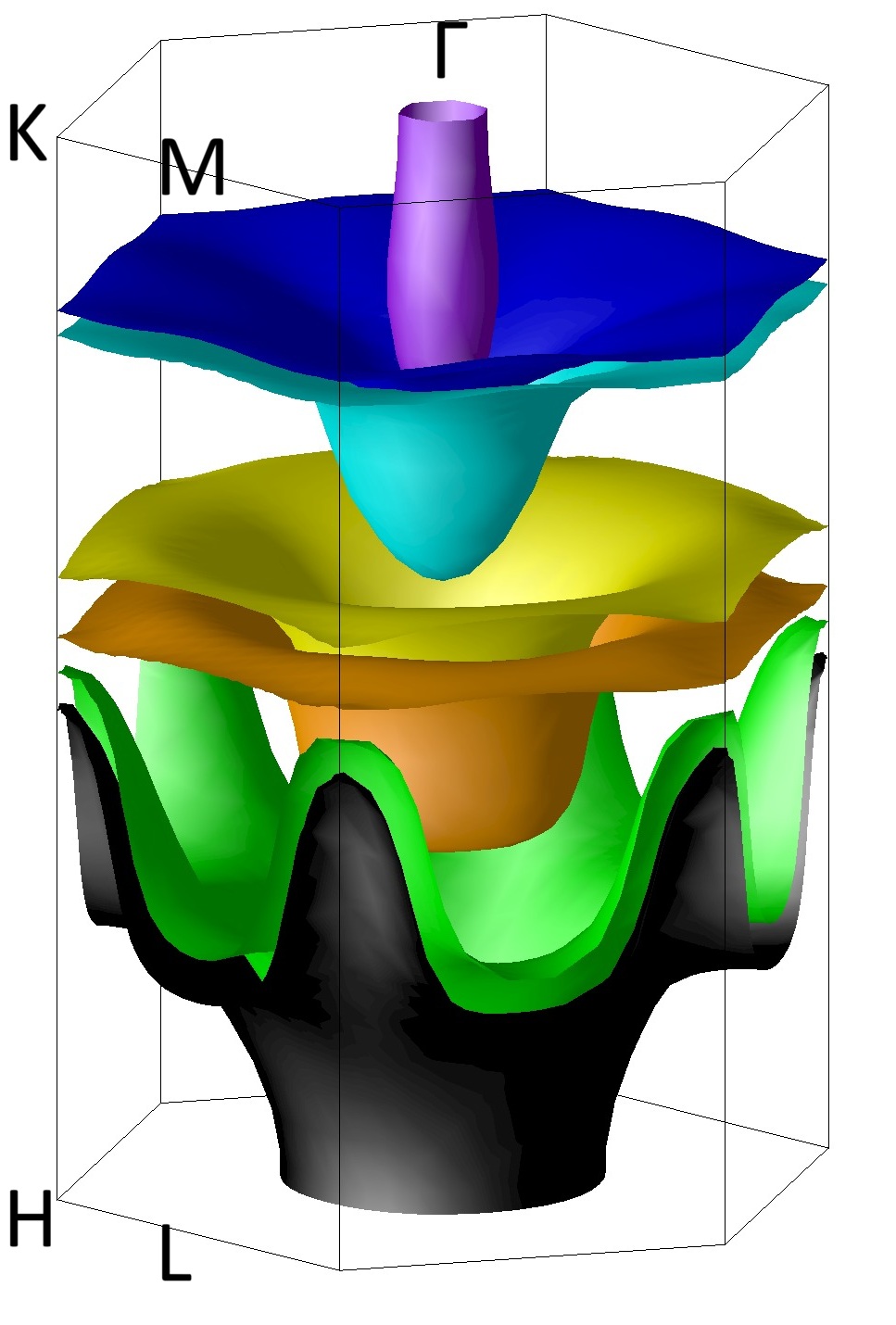}
\includegraphics[width=.37\linewidth]{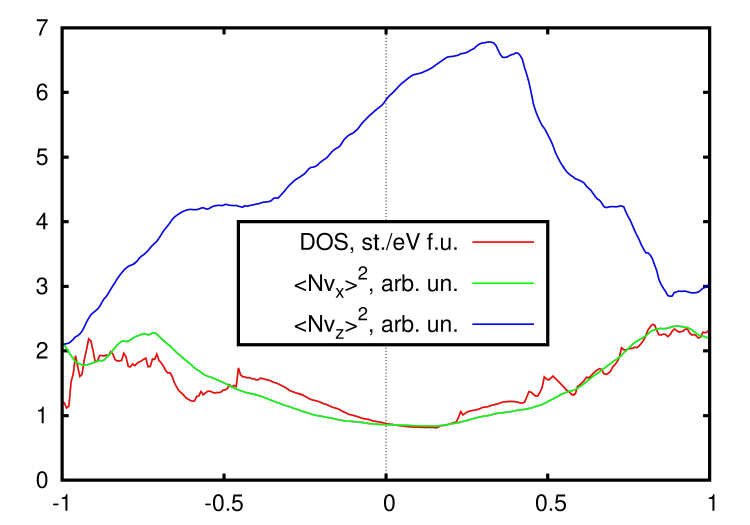}
  \caption{Left: Calculated DFT+U band structure, colored according the weight of the spin-up(down) contribution of Er states. Center: Fermi surface including spin-orbit coupling. Note large quasi-1D pieces of the  Fermi surfaces. The lower half of the Fermi surface is shown. Right: Density of states and the transport function $\langle Nv_i^2\rangle$.}
  \label{FS}
\end{figure*}

According to Hund’s rules, the Er-$4f$ shell carries a spin moment of $3~\mu_\text{B}$ and an orbital moment of $6~\mu_\text{B}$.
In our calculations, assuming the FM order, the Er-$4f$ shell has a total moment of $\sim9\mu_\text{B}$, with an additional small $5d$ contribution of $0.05 \mu_\text{B}$, while Pd atoms host on-site moments of
$0.1$–-$0.2\mu_\text{B}$ aligned antiparallel to Er.

Figure~\ref{fig:mae} shows the total energy $E(\theta,\phi)$ as a function of spin quantization direction, defined by polar and azimuthal angles $\theta$ and $\phi$, with $U=10$ eV applied
to the Er-$4f$ states.
The compound exhibits strong easy-axis anisotropy, with a uniaxial MAE of $\sim8.3$ meV/f.u., while the in-plane anisotropy is much smaller ($\sim0.5$ meV/f.u.), with the [210] direction being the hardest.
Varying $U$ between 8--14 eV changes the uniaxial MAE by about 15\% but has little effect on the total magnetization (See Table S2 and S3 in Supplementary).
The strong easy-axis anisotropy is consistent with the experiments in the present work and with the pronounced easy-plane anisotropy observed in HoAgGe \cite{bhandari2025tunable}.
Since Ho and Er are particle–hole symmetric with respect to the $4f$ charge asphericity  of their fully-polarized Hund's-rule ground states, the crystal-field model predicts opposite $E(\theta,\phi)$ trends under a constant
potential~\cite{lee2025ncm}.

In Fig. \ref{FS} we show the calculated bands and Fermi surface using $U=9$ eV, $J=1$ eV, including spin-orbit coupling and assuming a ferromagnetic arrangement. The band structure shows that the system is metallic with several bands crossing the Fermi energy $E_F$. 
Bands near $E_F$ exhibit some spin dependence as indicated by the color scale representing spin-up/down Er weight. 

Both exchange and spin-orbit coupling (SOC) contribute to the observed band splitting.
The relatively more uniform spin splitting throughout the band plot has the exchange origin and arises predominantly from Er 5d states hybridized with Pd and Pb, as the Er 4f states are well-removed from the Fermi level.
Near the $\Gamma$ point, however, the contribution of SOC becomes significant.
In supplementary materials, we also verified that in the nonmagnetic state without SOC, there are initially four bands degenerate 
at the Fermi level. SOC splits them into two pairs of spin-degenerate bands (upper and lower). The remaining spin degeneracy within each pair is then lifted by exchange splitting. The upper pair of bands, situated immediately above and below the Fermi level, presents an interesting case.

The dispersion is highly anisotropic, with flatter bands along certain directions, indicating strongly direction-dependent effective masses. Close inspection of the band structure near the $\Gamma$ point reveals a perfect saddle point (hole-like in-plane, electron-like out of plane), which in Fig. \ref{FS}(left) is spin-split because of the assumed ferromagnetic order (SOC alone does not introduce any splitting of the saddle point as shown in Fig. S2 (b). Therefore, we expect that in an antiferro- or paramagnetic state, the saddle point would appear exactly at the Fermi level, within the accuracy of the calculations.

The middle panel of Fig. \ref{FS} shows the calculated Fermi surface, which features an extended tubular sheet along $k_z$ and perpendicular flat  segments. 
The former is the outer sheet of the spin-split saddle point Fermi surfaces (the inner sheet is not visible). Remember that a saddle point situated exactly at the Fermi level results in a conical Fermi surface with two cones connected at the vertices.  The flat segments arise from weakly dispersive bands near the Fermi energy. Such quasi-one-dimensional (quasi-1D) Fermi surfaces indicate strong electronic anisotropy. The calculated Drude transport function, i.e. Fermi velocity squared averaged over the Fermi surface, is shown in the right panel of Fig. \ref{FS}. The calculated transport anisotropy, $\langle Nv_{z}^2\rangle/\langle Nv_{x}^2\rangle \propto$ $\sigma_{z}$/$\sigma_{x}$, is 6.6, in good agreement with the experimental data in Fig. \ref{Fig4}. 

\subsection{Conclusion}
In summary, we have successfully grown single crystals of ErPdPb, which crystallizes in the noncentrosymmetric ZrNiAl-type structure, and conducted a comprehensive investigation of its physical properties, including magnetic susceptibility, resistivity, magnetoresistance, and heat capacity, complemented by first-principles calculations. Our measurements reveal that ErPdPb undergoes easy-axis antiferromagnetic ordering below 2.2 K, with clear signatures observed across magnetic, transport, and thermodynamic probes. The observation of 1/3 magnetization plateau, together with a large frustration parameter, suggests that the Er spins, which reside on slightly distorted kagome planes, experience significant magnetic frustration.

Interestingly, the magnetization tends toward saturation at high fields without exhibiting discrete magnetization plateaus, even well above the Néel temperature (up to 15 K), indicating the presence of persistent spin fluctuations. Moreover, a particularly notable feature emerges in the heat capacity measurements: below approximately 0.5 K, the heat capacity increases as the temperature decreases, down to the lowest measured temperature of 50 mK. This behavior could signal either the onset of Kondo effect or a secondary magnetic phase transition below 50 mK.

The calculated electronic structure reveals a quasi-1D Fermi surface, consistent with the experimentally observed transport anisotropy. The band structure also contains a saddle point very close to the Fermi energy, a feature that can place the system near Fermi surface instabilities. Taken together, these characteristics mark ErPdPb as an intriguing materials platform: a frustrated antiferromagnet in a noncentrosymmetric structure, at the verge of quantum criticality, with magnetic ordering occurring at ultra-low temperatures. These findings motivate further exploration, particularly through chemical substitution or applied pressure, to suppress the ordering temperature and investigate emergent phenomena such as unconventional superconductivity or non-Fermi-liquid behavior in the vicinity of a quantum critical point.

\begin{acknowledgments}
This work was primarily supported by the US Department of Energy, Office of Science, Basic Energy Sciences, Materials Science and Engineering Division. IM was supported by the Office of Naval Research through grant \#N00014-23-1-2480. Work at Ames National Laboratory
is supported by the U.S. Department of Energy, Office of Basic Energy Sciences, Materials Sciences and Engineering Division. LK acknowledges Research Computing at the University of Virginia for providing computational resources for a part of this research. Ames National Laboratory is operated for the U.S. Department of Energy
by Iowa State University under Contract No. DE-AC02-07CH11358.
\end{acknowledgments}


%

\pagebreak
\maketitle
\section{S1: Structural characterization}

\begin{figure*}[ht!]
\begin{center}
\includegraphics[width=.8\linewidth]{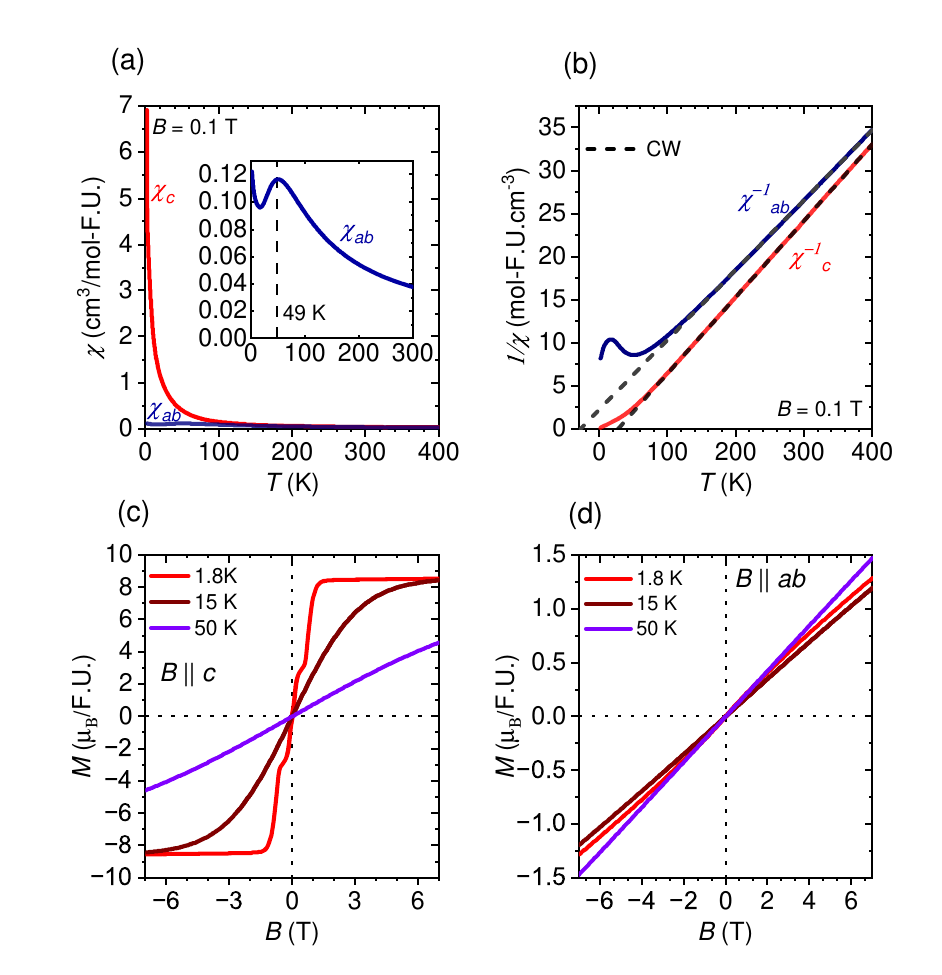}
     \caption{\small Magnetic properties of ErPdPb  measured on a second sample from a different growth batch than that used to measure the properties reported in Fig. 3 of the main text. (a) Magnetic susceptibility measured in a magnetic field of 0.1 T applied along crystallographic $c$-axis ($\chi_c$) and within the $ab$-plane ($\chi_{ab}$) measured with a field-cooled (FC) protocol. Inset highlights a broad feature centered at 49 K for $\chi_{ab}$ (b)  Inverse susceptibility for $B$ along $c$-axis  ($\chi^{-1}_{c}$) and along $ab$-plane ($\chi^{-1}_{ab}$). (c) Magnetic moment for $B||c$-axis. (d) Magnetic moment for $B||ab$-plane.} 
      \label{FigS1}
    \end{center}
\end{figure*}
\begin{table}[H]
\caption{Anisotropic thermal Parameters (Å$^2$×10$^2$) for ErPdPb. Numbers in parentheses are estimated standard deviation.}
\label{T2}
\centering
\begin{tabular}{c@{\hspace{0.3cm}}c@{\hspace{0.3cm}}c@{\hspace{0.3cm}}c@{\hspace{0.3cm}}c@{\hspace{0.3cm}}c@{\hspace{0.3cm}}c}
\hline
Atom & U$_{11}$ & U$_{22}$ & U$_{33}$ & U$_{23}$ & U$_{13}$ & U$_{12}$ \\
\hline
Er1 & 0.0126(4) & 0.0134(5) & 0.0108(5) & 0 & 0 & 0.0067(3) \\
Pd1 & 0.017(1)  & 0.017(1)  & 0.024(2)  & 0 & 0 & 0.0085(5) \\
Pd2 & 0.0111(7) & 0.0111(7) & 0.021(1)  & 0 & 0 & 0.0055(3) \\
Pb1 & 0.0127(4) & 0.0120(5) & 0.0159(5) & 0 & 0 & 0.0060(3) \\
\hline
\end{tabular}
\end{table}


\section{S2: Magnetic properties reverification}

Fig.\ref {FigS1} here shows magentic measurments of a single crystal of ErPdPb from another batch for reproducibility which is consistent to the plots included in the text.  The inverse susceptibility Curie-Weiss (CW) fit to $\chi^{-1}_{c}$ reveals $\mu_{\mathrm{eff}} = 9.5$ $\mu_B$ and $\theta_{\mathrm{CW}} = 27.3$ K. While fit for $\chi^{-1}_{ab}$ reveals $\mu_{\mathrm{eff}} = 9.9$ $\mu_B$ and $\theta_{\mathrm{CW}} = -28.1$ K. The magentic moment along $B||c$-axis has 1/3 steps below 3K (Fig.\ref {FigS1}c) while along $B||ab$-plane it is linear above 3K (Fig.\ref {FigS1}d)

\section{S3: First principles calculations}
\begin{figure*}[ht!]
\begin{center}
\includegraphics[width=.6\linewidth]{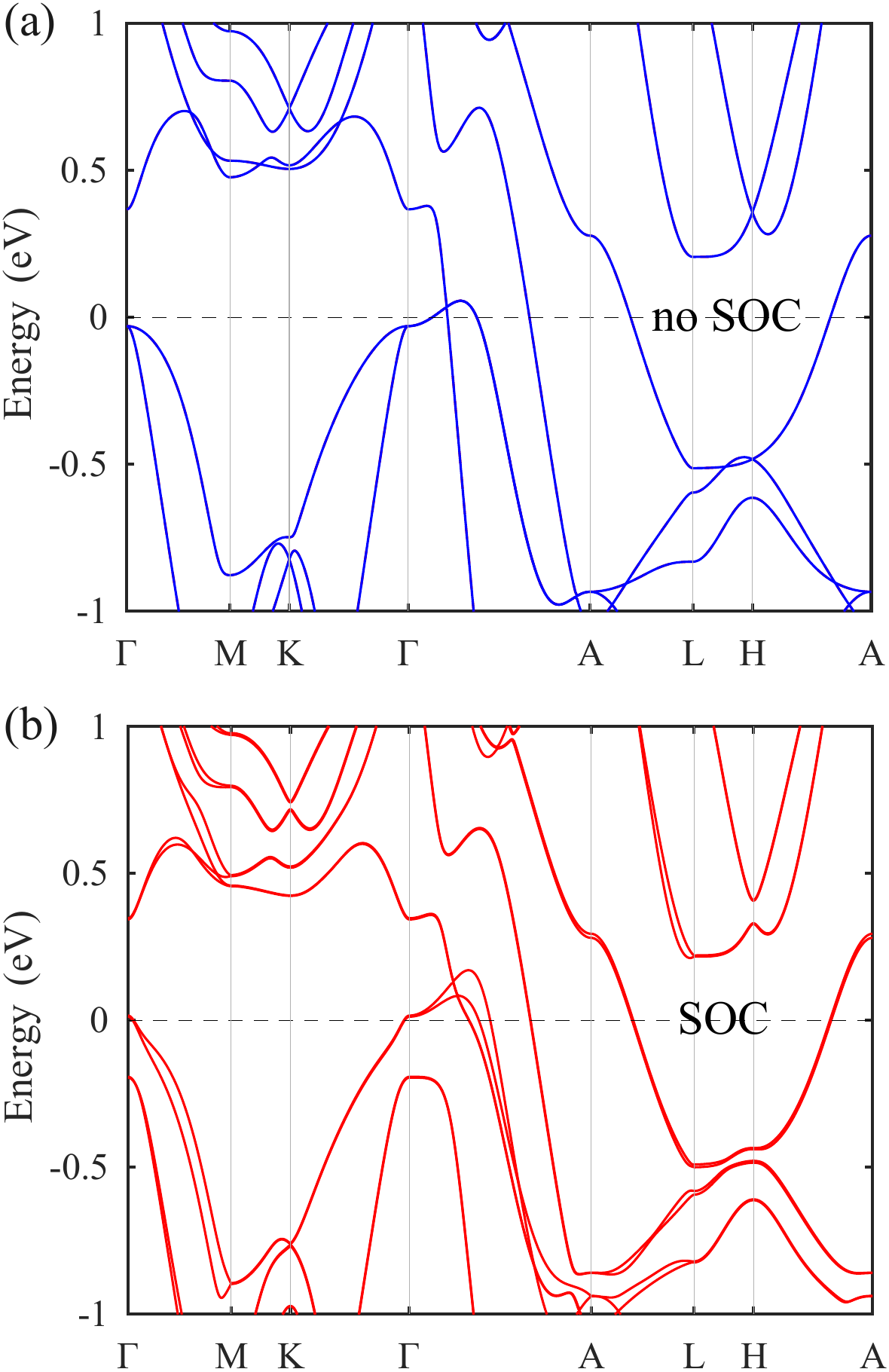}
    \caption{\small (a) Electronic band structure calculated in non-magnetic state without SOC (b)  Electronic band structure calculated in non-magnetic state with SOC.}
      \label{SOCandWOSOC}
    \end{center}
\end{figure*}

\begin{table}[htbp]
	\caption{Calculated Total magnetic moment $M$, on-site Er, Pd magnetic moments, $m_\text{Er}$ and $m_\text{Pd}$ with various $U$ values applied on Er-$4f$ orbitals, spin contribution $m_\text{Er}^s$, the Er-$5d$ spin contribution ($m_{5d}^s$) and the Er-$4f$ spin contribution ($m_{4f}^s$)and orbital ($m_{4f}^l$).} 
	\label{tbl:mag}
	\bgroup
	\def\arraystretch{1.2}                                
	\begin{tabular*}{\linewidth}{c@{\extracolsep{\fill}}ccccccc}
		\hline \hline  
		$U$ &   $m_{4f}^l$  & $m_{4f}^s$&$m_{5d}^s$ & $m_\text{Er}^s$ & $m_\text{Er}$ &  $m_\text{Pd}$ &  $M$  \\ \hline
		8   &    5.93       & 2.97      &0.04        & 3.03            & 8.96          & -0.012          &  8.97 \\
		10  &    5.95       & 2.98      &0.04        & 3.04            & 8.99          & -0.016          &  9.00 \\
		12  &    5.96       & 2.98      &0.04        & 3.04            & 9.00          & -0.018          &  9.01 \\
		14  &    5.97       & 2.99      &0.04        & 3.04            & 9.01          & -0.020          &  9.02 \\ \hline
	\end{tabular*}
	\egroup	
\end{table}

Table \ref{tbl:mag} depicts the total magnetic moment $M$, the magnetic moments of Er, Pd, $m_\text{Er}$ and $m_\text{Pd}$, in ErPbPd calculated with various values  of $U$ applied on the Er orbitals $4f$. Er magnetic moment is further resolved into its spin contribution $m_\text{Er}^s$, the contributions from Er-$5d$ spin ($m_{5d}^s$), and the contributions from Er-$4f$ spin ($m_{4f}^s$)and orbital ($m_{4f}^l$). The total magnetization $M$ is in the unit of $\mu_B$/f.u., while all other components are in the unit of $\mu_B$/atom and calculated inside the muffin-tin (MT) spheres.The magnetic moments of Pd and Pb are negligible, although the magnetic moment of Pd is listed. According to Hund's rules, Er-$4f$ is expected to have $S=\frac{3}{2}$, $L=6$, and $J=\frac{15}{2}$.

\begin{table}[htbp]
	\caption{
	The $U$-dependent magnetic anisotropy energy (MAE) (in meV/f.u.) evaluated for total energies along spin orientations [100], [210], and [001]
}
                     
	\label{tbl:mag2}
	\bgroup
	\def\arraystretch{1.2}                                
	\begin{tabular*}{\linewidth}{c@{\extracolsep{\fill}}cccccc}
		\hline \hline  
		$U$ (eV)  & $E_{100}$ & $E_{210}$ &   $E_{001}$ \\ \hline
		8    & 8.74      & 9.40     &    0        \\
		10   & 8.32      & 8.86     &    0        \\
		12   & 7.86      & 8.25     &    0        \\
		14   & 7.42      & 7.69     &    0        \\ \hline
	\end{tabular*}
	\egroup	
\end{table}

Table \ref{tbl:mag2} depicts the $U$-dependent magnetic anisotropy energy (MAE) (in meV/f.u.) which is evaluated by calculating the total energies for spin orientations along the [100], [210], and [001] directions, using $U$ values of 8, 10, 12, and 14 eV. The results indicate that $\erpbpd$ exhibits uniaxial magnetic anisotropy (MA), and the overall MAE decreases with increasing $U$. At $U = 8$ eV, the in-plane MAE is approximately 8\% of the out-of-plane MAE , and this ratio decreases to about 3\% as $U$ increases to 14 eV.

\end{document}